\documentclass[12pt]{article}

\usepackage{amssymb}
\usepackage{amsmath}
\usepackage{amsbsy}
\usepackage{bm}
\usepackage{graphicx}
\usepackage{mathrsfs}

\usepackage{color}
\usepackage{doi}
\renewcommand{\doiurl}{https://doi.org/}
\renewcommand{\doitext}{https://doi.org/} 
\usepackage[square,comma,numbers,sort&compress]{natbib}

\def\bea{\begin{align}}
\def\bean{\begin{align*}}

\def\intl{\int\limits}

\def\eps{\varepsilon}

\begin{document}
\title{Fugacity versus chemical potential in nonadditive generalizations of the ideal Fermi-gas}
\author{Andrij Rovenchak, Bohdana Sobko\\
{\it Department for Theoretical Physics,}\\
{\it Ivan Franko National University of Lviv,}\\
{\it 12 Drahomanov St., Lviv, UA--79005, Ukraine}}

\maketitle

\abstract{
We compare two approaches to the generalization of the ordinary Fermi-statistics based on the nonadditive Tsallis $q$-exponential used in the Gibbs factor instead of the conventional exponential function. Both numerical and analytical calculations are made for the chemical potential, fugacity, energy, and the specific heat of the ideal gas obeying such generalized types of statistics. In the approach based on the Gibbs factor  containing the chemical potential, high temperature behavior of the specific heat significantly deviates from the expected classical limit, while at low temperatures it resembles that of the ordinary ideal Fermi-gas. On the contrary, when the fugacity enters as a multiplier at the Gibbs factor, the high-temperature limit reproduces the classical ideal gas correctly. At low temperatures, however, some interesting results are observed, corresponding to non-zero specific heat at the absolute zero temperature or a finite (non-zero) minimal temperature. These results, though exotic from the first glance, might be applicable in effective modeling of physical phenomena in various domains.

}

\section{Introduction}
Nonextensive and nonadditive generalizations of entropy originated in the information theory \cite{Renyi:1961,Daroczy:1970} and were introduced in physical problems by Tsallis \cite{Tsallis:1988}. The approaches based on Tsallis's generalization are applicable to problems, where, for instance, long-range interactions or non-Markovian memory effects are essential \cite{Abe&Okamoto:2001}. These include both physical phenomena \cite{Marques_etal:2015,Tripathy_etal:2016,Kusakabe_etal:2019} and interdisciplinary applications \cite{Gell-Mann&Tsallis:2004,Pavlos_etal:2015,Arunkumar_etal:2016,Ruiz&deMarcos:2018,Neuman_etal:2018,Rovenchak&Buk:2018}.

Various formulations of nonadditive generalizations are known for quantum Bose- and Fermi-distributions \cite{Buyukkilic_etal:1995,Pennini_etal:1996,Tanatar:2002,Ou&Chen:2003,Aragao-Rego_etal:2003,Martinez_etal:2004,Mohammadzadeh_etal:2016,Chung&Algin:2017}. Note that the Bose-systems are studied to a larger extent, perhaps due to the fascinating Bose-condensation phenomenon \cite{Algin&Okun:2017,Adli_etal:2019}. While the deformations of the Fermi-statistics are often studied alongside the Bose-statistics, in some works deformed fermions are a sole subject of analysis \cite{Pennini_etal:1996,Martinez_etal:2004,Rozynek:2015}.

In the present paper, we use a phenomenological model previously studied for nonadditive generalizations of the fractional Polychronakos statistics \cite{Rovenchak:2014PRA} and the ideal Bose-gas \cite{Rovenchak:2018LTP}.
To obtain thermodynamic properties as functions of temperature $T$, we apply the standard procedure linking the number of particles $N$ in a system with chemical potential $\mu$ or fugacity $z=e^{\mu/T}$:
\bea\label{eq:N}
N = \sum_j n(\eps_j,z,T) = \intl_0^\infty d\eps\; g(\eps)n(\eps,z,T),
\end{align}
where $n(\eps,z,T)$ are mean occupation numbers, given in the conventional Fermi-statistics by
\bea\label{eq:Fermi}
n(\eps,z,T) = \frac{1}{z^{-1}e^{\eps/T} + 1}.
\end{align}
The density of states $g(\eps)$ is introduced for convenience to substitute the summation over all levels $\eps_j$ with integration over energies $\eps$. To cover a vast diversity of problems, the density of states can be written as
\bea\label{eq:g}
g(\eps) = NA\,\eps^{s-1},
\end{align}
where, for instance, $s=D/2$ for free particles in a $D$-dimensional space, $s=D$ for $D$-dimensional harmonic oscillators oscillators, etc. Note that for convergence of the integral in Eq.~(\ref{eq:N}) we must require that $s>1$. The factor $A$ is a constant independent of  energy and is defined by such parameters as concentration of particles, mass of particles, harmonic oscillator frequencies, etc., depending of the specific system under consideration.

With the chemical potential or fugacity as functions of temperature, we can calculate the total energy
\bea\label{eq:E}
E = \sum_j \eps_j n(\eps_j,z,T) = 
\intl_0^\infty d\eps\; \eps g(\eps)n(\eps,z,T)
\end{align}
and the isochoric heat capacity
\bea\label{eq:CV}
C_V = \left(\frac{\partial E}{\partial T}\right)_V.
\end{align}

The abovementioned functions, $\mu(T)$, $z(T)$, $E(T)$, and $C_V(T)$ are in the focus of the present paper. In the following sections, we apply two different procedures to introduce nonadditive generalizations of the Fermi-distribution (\ref{eq:Fermi}), provide numerical results for the whole temperature range and then analyze low- and high-temperature regimes analytically.

\section{Deforming the Gibbs factor}
We analyze two approaches to the generalization of the Fermi-distribution
%\bea
%n^{\rm Fermi}(\eps,\mu,T) = \frac{1}{e^{(\eps-\mu)/T} + 1}
%\end{align}
using the nonadditive Tsallis $q$-exponential \cite{Tsallis:1994}
\bea\label{eq:qexpTsallis-def}
e_q^x =\! \left\{\!\!
\begin{array}{ll}
e^x &\!\!\!\textrm{for\ } q=1,\\[6pt]
[1+(1-q)x]^{1/(1-q)} &\!\!\!\textrm{for\ }q\neq1\ \textrm{and}\ 
1+(1-q)x > 0,\\[6pt]
0^{1/(1-q)}  &\!\!\!\textrm{for\ }q\neq1\ \textrm{and}\ 
1+(1-q)x \leq 0.
\end{array}
\right.\hspace*{-1em}
\end{align}
As the first modification, we consider the following substitution of the Gibbs factor $e^{(\eps-\mu)/T}$ in the Fermi-distribution: 
\bea
n_1(\eps,\mu,T) = \frac{1}{e_q^{(\eps-\mu)/T} + 1}.
\end{align}
The second modification is very similar, namely
\bea
n_2(\eps,z,T) = \frac{1}{z^{-1}e_q^{\eps/T} + 1}.
\end{align}

\begin{figure}[h]
\centerline{%
\includegraphics[width=0.64\textwidth]{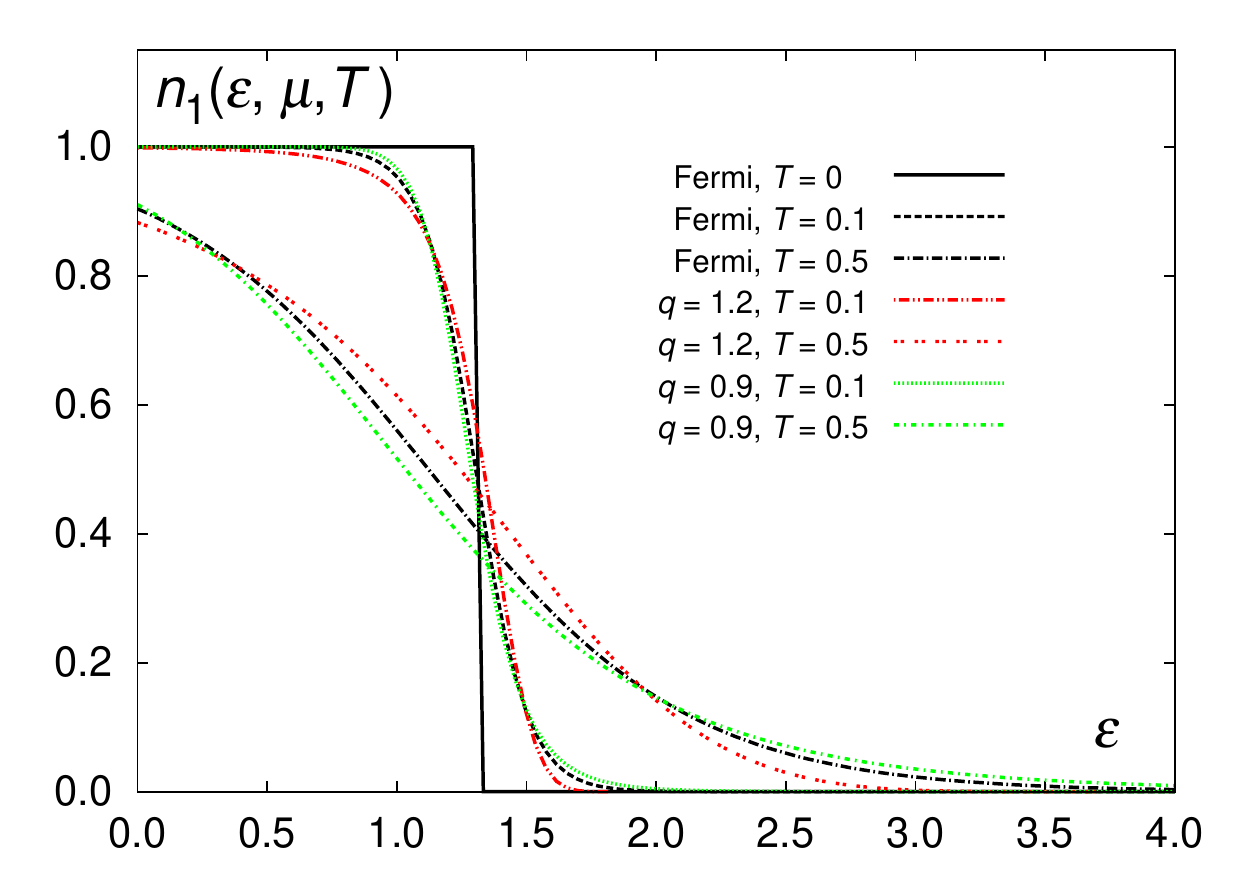}}%
\bigskip
\centerline{%
\includegraphics[width=0.64\textwidth]{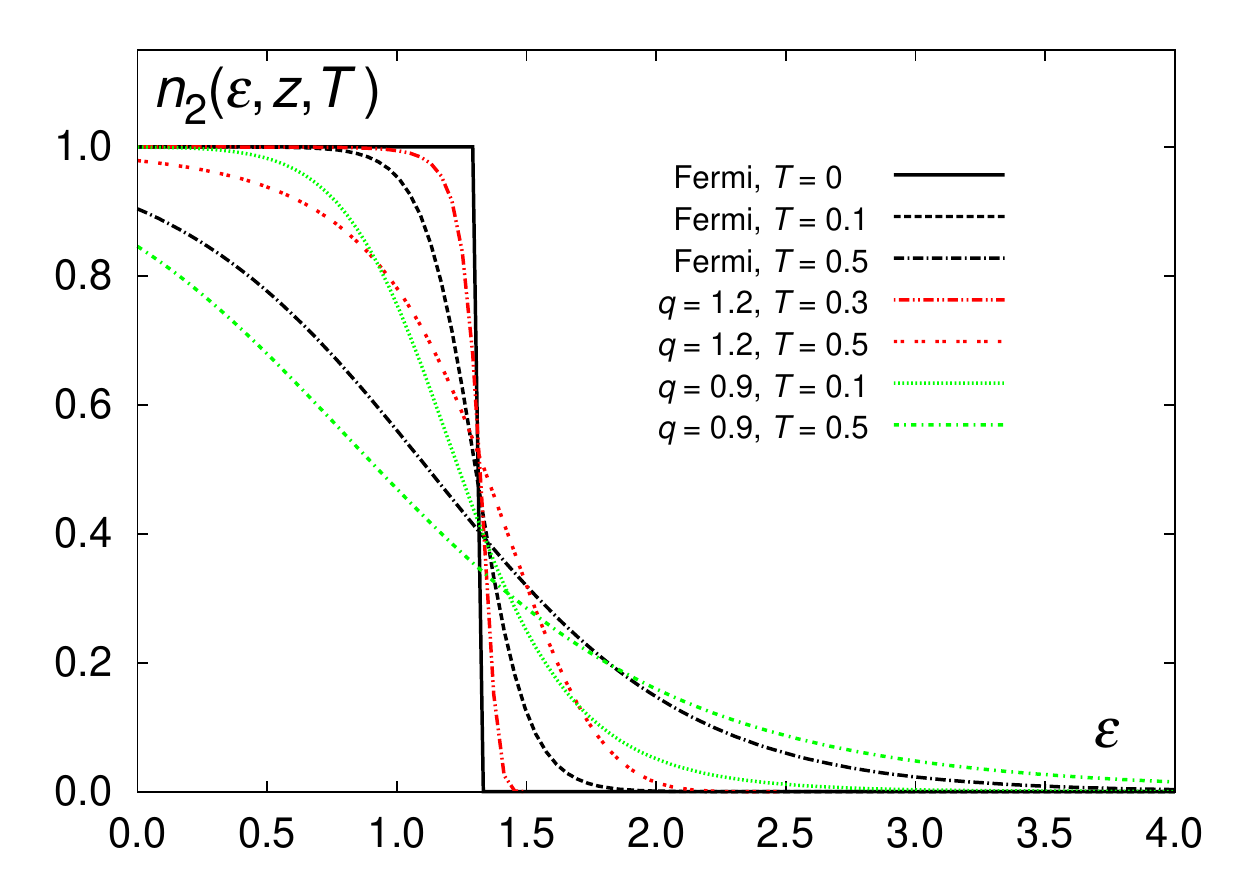}
}
\caption{Occupation numbers $n_1$ and $n_2$ at different values of $q$ and temperature $T$. The chemical potential and fugacity are fixed by Eq.~(\ref{eq:N}).}
\label{fig:n-eps}
\end{figure}

\clearpage 

Since the factorization rule breaks for the $q$-exponentials, $e_q^{x+y}\neq e_q^x e_q^y$, one cannot establish a simple connection  between the chemical potential and fugacity similar to $z=e^{\mu/T}$ for the ordinary case. Moreover, even if we can introduce a deformed fugacity as $z_q=e_q^{\mu/T}$, its inverse would not be related to the chemical potential in a simple manner either, $z_q^{-1}\neq e_q^{-\mu/T}$. To be precise, the following relations hold for the Tsallis $q$-exponentials \cite{Yamano:2002}: 
\bea\label{eq:qexp-rules}
(e_q^x e_q^y)^{1-q} = (e_q^{x+y})^{1-q} + (1 - q)^2xy,
\qquad
\left[e^x_q\right]^{-1} = e^{-x}_{2-q}.
\end{align}

In Figure~\ref{fig:n-eps} the functions $n_1$ and $n_2$ are shown for different values of $q$ and temperature $T$. Here and further, for numerical calculations we set in the density of states (\ref{eq:g}) constant $A=1$ and power $s=3/2$.

In the case of the modification $n_1$ involving the chemical potential, it is quite straightforward to demonstrate that the limit of $T\to0$ yields a standard Fermi step. Indeed, for $q<1$, the expression $[1+(1-q)(\eps-\mu)/T]$ is negative for $\eps<\mu$ yielding zero when raised to the positive power $1/(1-q)$, according to Eq.~(\ref{eq:qexpTsallis-def}). And vice versa, for $q>1$ we have a positive infinity to the negative power, hence zero as well. A direct analysis of the second modification, $n_2$ involving fugacity, is a bit more complicated and we will postpone it for subsequent sections.

\section{Numerical results}
\begin{figure}[h]
\centerline{%
\includegraphics[width=0.8\textwidth]{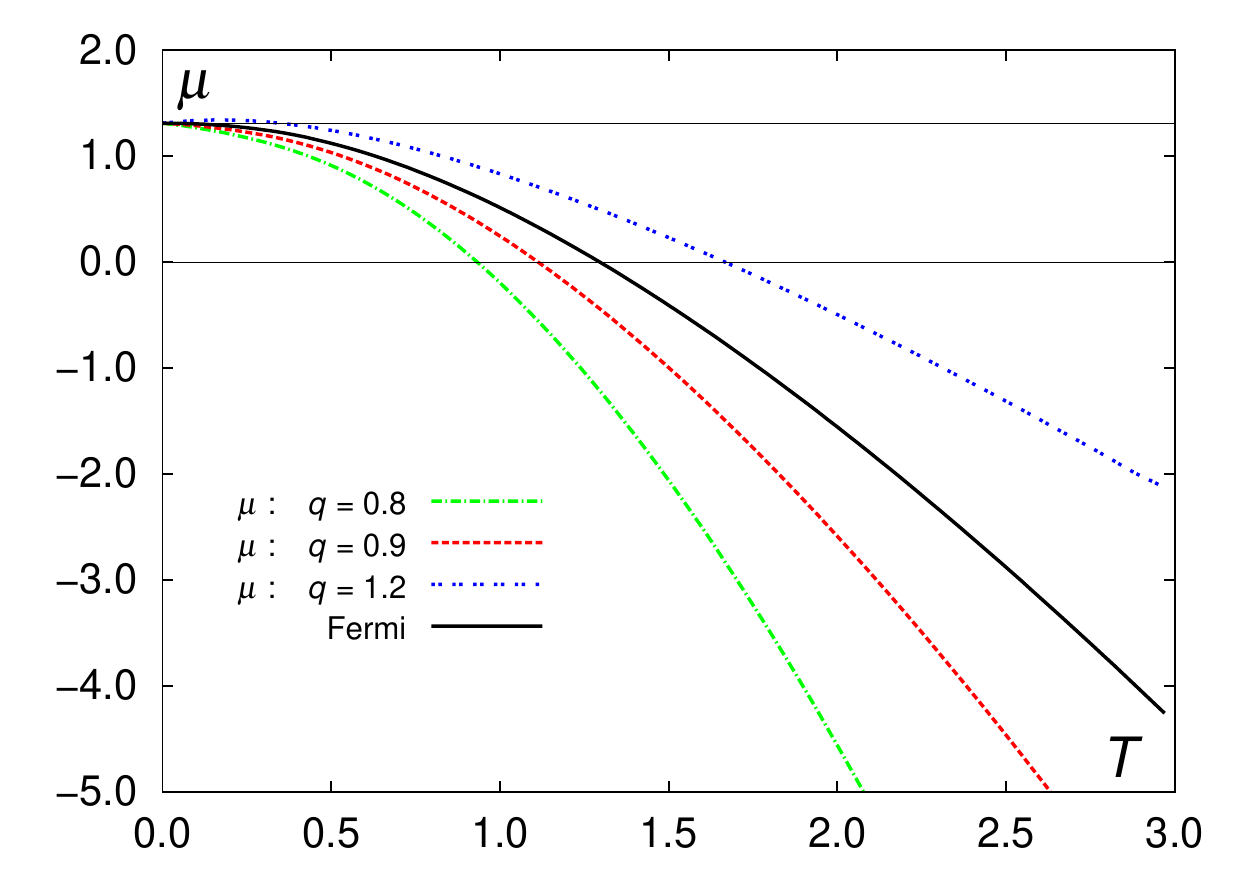}
}
\caption{Chemical potential as a function of temperature $T$ at different values of $q$ for the first model.}
\label{fig:mu}
\end{figure}

\begin{figure}[h]
\centerline{%
\includegraphics[width=0.8\textwidth]{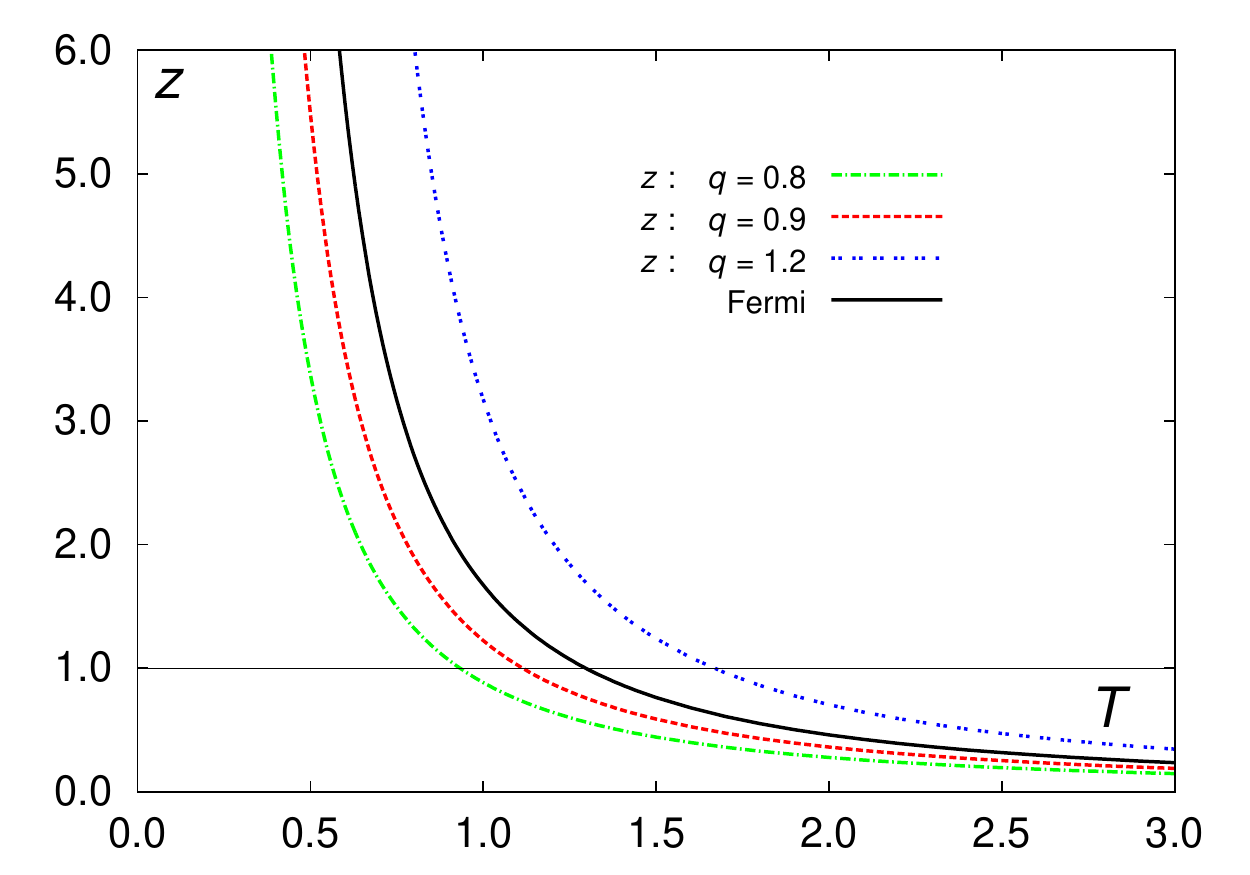}
}
\caption{Fugacity as a function of temperature $T$ at different values of $q$ for the second model.}
\label{fig:z}
\end{figure}

The temperature dependences of the chemical potential $\mu$ and fugacity $z$ calculated numerically are shown in Figs.~\ref{fig:mu} and \ref{fig:z}. They are by solving numerically the equation
\bea
N = \intl_0^\infty g(\eps)n_{1,2}(\eps,\cdot,T)\,d\eps,
\end{align}
where the dot stands for the chemical potential $\mu$ in the case of $n_1$ and for the fugacity $z$ in the case of $n_2$.

\clearpage

\begin{figure}[h]
\centerline{%
\includegraphics[width=0.8\textwidth]{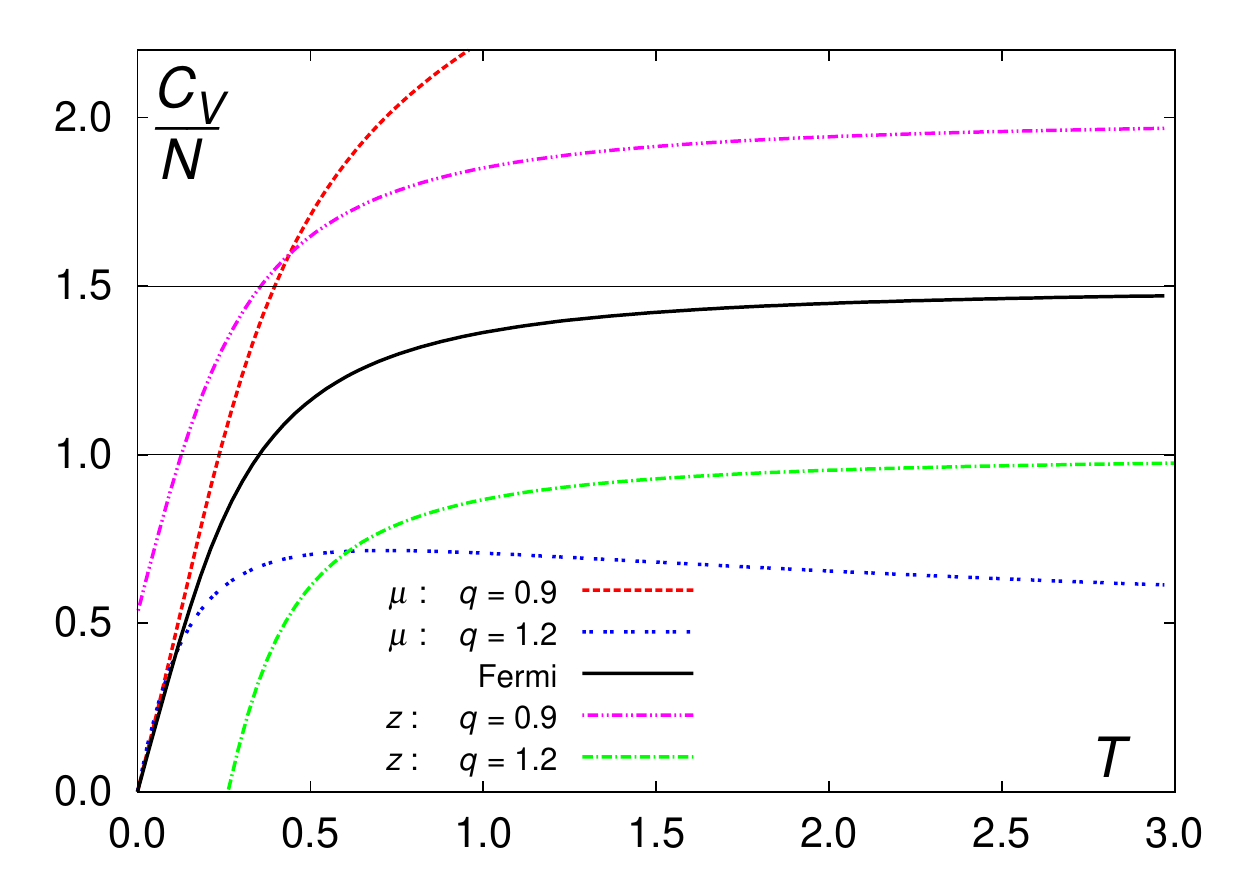}
}
\caption{Specific heat as a function of temperature $T$ at different values of $q$.
Models 1 and 2 demonstrate different behavior in low and high-temperature limits.}
\label{fig:CV}
\end{figure}

We then calculate energy
\bea
E = \intl_0^\infty \eps\,g(\eps)n_{1,2}(\eps,\cdot,T)\,d\eps
\end{align}
and the isochoric heat capacity (\ref{eq:CV}). The results for the isochoric specific heat $C_V/N$ are shown in Fig.~\ref{fig:CV}.

The results of numerical calculations give us some hints facilitating  analytical derivations presented in the next sections. In particular, the behavior of $\mu$ and $z$ for different values of the statistics parameter $q$ in Figs.~\ref{fig:mu} and \ref{fig:z} suggests that there is a smooth transition between the domains of $q<1$ and $q>1$ separated by the ordinary Fermi-gas corresponding to $q=1$.  

The most interesting, however, is the graph for the specific heat. One would expect that in the high-temperature limit the value of $C_V$ tends to a constant (equal to $sN$ in the case of an ordinary ideal gas). But the first model, where the modified Gibbs factor contains the chemical potential, the heat capacity seems to increase infinitely if $q<1$ and to tend to zero if $q>1$, as shown for the nonadditive ideal Bose-gas \cite{Rovenchak:2018LTP}. For the second model, with $z$, the classical limit is a constant depending on $q$.

The low-temperature domain has its own peculiarities. Namely, the temperature behavior of first model nearly coincides with that of the ordinary Fermi-gas, while for the second model two unexpected results are observed. For $q<1$, the value of the heat capacity does not approach zero at $T\to0$. For $q>1$, the heat capacity becomes zero at a finite (non-zero) temperature. We confirm these observations analytically in subsequent sections.

%\clearpage

\section{High temperatures and classical limit for the first model}
In \cite{Rovenchak:2018LTP} it was shown how to obtain the high-temperature behavior of the ideal Bose-gas satisfying the nonadditive modification of the Gibbs factor involving the chemical potential, with $e^{(\eps-\mu)/T}$ substituted by $e_q^{(\eps-\mu)/T}$. The same approach can be used for the respective Fermi-gas model.

From Fig.~\ref{fig:mu} we can see that, naturally, the ordinary Fermi-gas separates cases $q<1$ and $q>1$. For fermions, the chemical potential tends to the classical limit as $T\to\infty$,
\bea
\frac{\mu}{T} = -\frac{|\mu|}{T} \to -\infty.
\end{align}
So, 
\bea
N = NA\intl_0^\infty \frac{\eps^{s-1}\,d\eps}{e^{(\eps-\mu)/T}+1} \simeq
NAT^s e^{\mu/T}\intl_0^\infty dx\, x^{s-1} e^{-x}.
\end{align}
We have thus the high-temperature limit
\bea
\mu \simeq -sT\ln T.
\end{align}

Now we consider the deformed case for $q<1$. The number of particles equals
\bea
N = NA\intl_0^\infty \frac{\eps^{s-1}\,d\eps}{e_q^{(\eps-\mu)/T}+1} \simeq
NAT^s \intl_0^\infty dx\, x^{s-1} \left[e_q^{x+|\mu|/T}\right]^{-1}
\end{align}
since the unity in the denominator is small comparing to the $q$-exponential. In the same fashion, the energy yields
\bea
E = NA\intl_0^\infty \frac{\eps^{s}\,d\eps}{e_q^{(\eps-\mu)/T}+1} \simeq
NAT^{s+1} \intl_0^\infty dx\, x^{s} \left[e_q^{x+|\mu|/T}\right]^{-1}.
\end{align}
After simple calculations \cite{Rovenchak:2018LTP} we arrive at the following high-temperature behavior for the chemical potential 
\bea
\mu = 
-\left[
A(1-q)^{\frac{1}{q-1}}\textrm{B}\left(s,\frac{1}{1-q}-s\right)
\right]^{\frac{q-1}{1+s(q-1)}} T^{\frac{1}{1+s(q-1)}},
\end{align}
where $\textrm{B}(x,y)$ is Euler's beta-function,
energy
\bea
\frac{E}{N} = 
\frac{\textrm{B}\left(s+1,\frac{1}{1-q}-s-1\right)}%
{\textrm{B}\left(s,\frac{1}{1-q}-s\right)}\,|\mu|,
\end{align}
and, respectively, the specific heat:
\bea\label{eq:CV-gamma}
\frac{C_V}{N} \propto T^{\frac{(1-q)s}{1+s(q-1)}} = T^{\gamma}.
\end{align}

The case of $q>1$ is a bit more tricky. The definition of the $q$-exponential (\ref{eq:qexpTsallis-def}) implies a finite upper limit of integration,
\bea
N = NAT^s\intl_0^{\frac{1}{q-1}-\frac{|\mu|}{T}}
\frac{x^{s-1}\,dx}{\left[1+(1-q)\left(x-\frac{|\mu|}{T}\right)\right]^{1/(1-q)}+1}
\end{align}
and the condition $\displaystyle{\frac{|\mu|}{T}<\frac{1}{q-1}}$. Setting
\bea
\frac{|\mu|}{T}=\frac{1}{q-1}-\delta,
\end{align}
where $\delta$ is a small number tending to zero at large temperatures, we can apply the procedure similar to the one described above for the $q<1$ case and finally obtain
\bea
\mu = \frac{1}{1-q}+
\left[
A(q-1)^{\frac{1}{q-1}}\textrm{B}\left(s,\frac{1}{q-1}+1\right)
\right]^{\frac{q-1}{1+s(q-1)}} T^{\frac{1}{1+s(q-1)}}
\end{align}
yielding the same temperature dependence for the specific heat as in (\ref{eq:CV-gamma}). Obviously, the ordinary exponential corresponding to $q=1$ ensures $C_V={\rm const}$ as $T\to \infty$.

\begin{figure}[h]
\centerline{%
\includegraphics[width=0.8\textwidth]{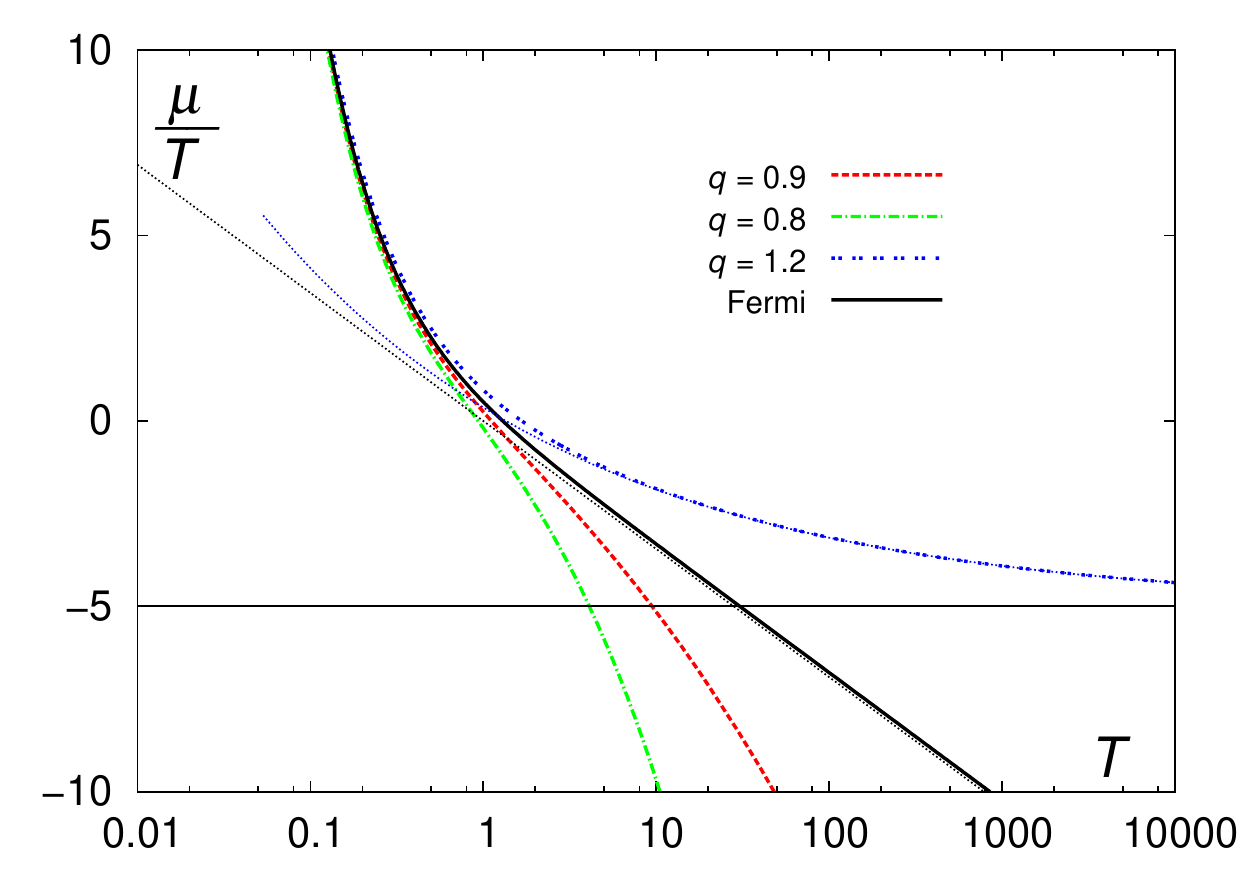}
}
\caption{Chemical potential as a function of temperature $T$ at different values of $q$ at high temperatures. Dotted lines show the asymptotic behavior.
}
\label{fig:muT-log}
\end{figure}

\begin{figure}[h]
\centerline{%
\includegraphics[width=0.8\textwidth]{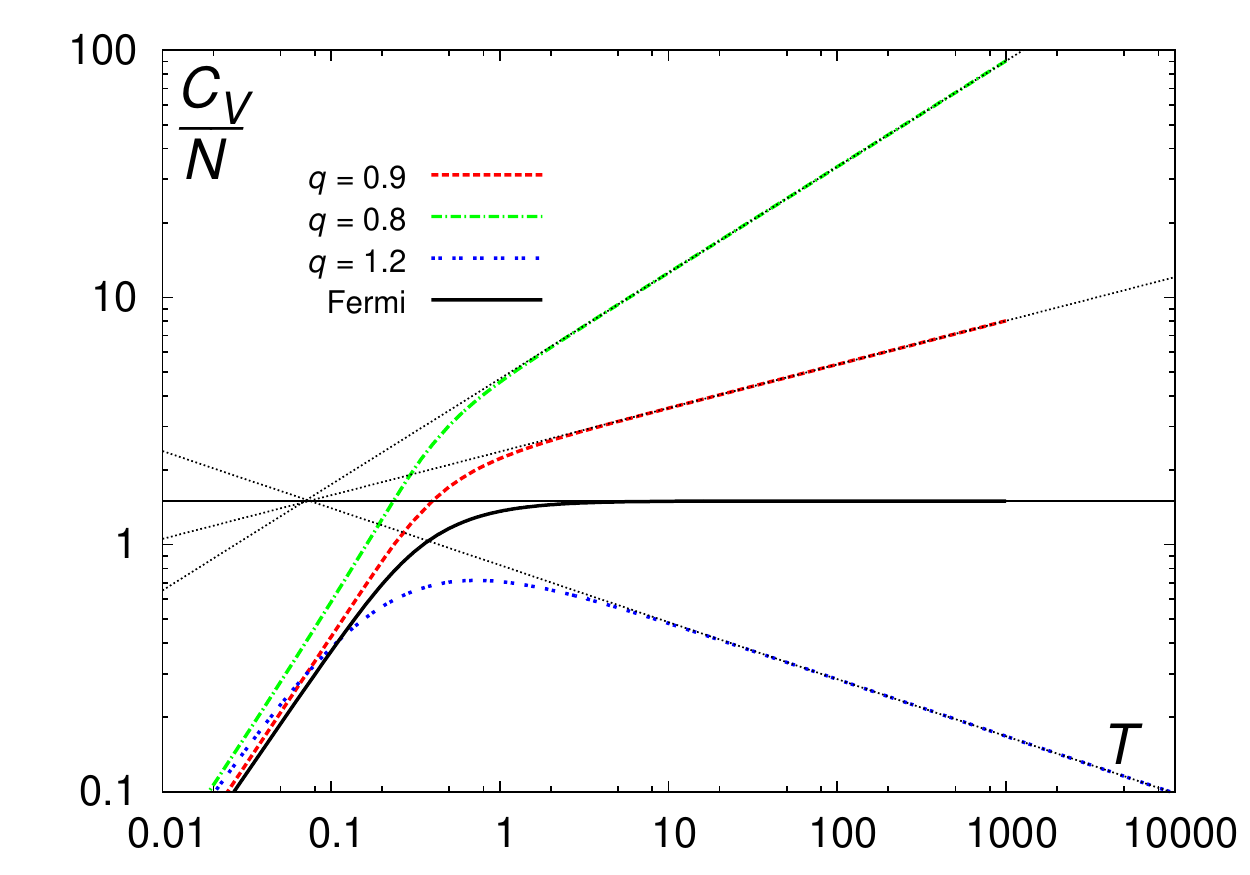}
}
\caption{Specific heat as a function of temperature $T$ at different values of $q$ at high temperatures. Dotted lines show the asymptotic behavior.
}
\label{fig:CV-log}
\end{figure}

The respective temperature behavior at high temperatures is shown in Figs.~\ref{fig:muT-log} and \ref{fig:CV-log}.

\clearpage

\section{High temperatures and classical limit for the second model}

For the model involving fugacity, the same analysis can be made as for the nonadditive Bose gas \cite{Rovenchak:2018LTP}. Since $z\to 0$ at high temperatures, the number of particles and energy are written as
\bea\label{eq:Nz0}
N = NA\intl_0^\infty \frac{\eps^{s-1}\,d\eps}{z^{-1}e_q^{\eps/T}+1} \simeq
zNAT^s \intl_0^\infty dx\, x^{s-1} \left[e_q^{x}\right]^{-1}
\end{align}
and
\bea\label{eq:Ez0}
E = NA\intl_0^\infty \frac{\eps^{s}\,d\eps}{z^{-1}e_q^{\eps/T}+1} \simeq
zNAT^{s+1} \intl_0^\infty dx\, x^{s} \left[e_q^{x}\right]^{-1}.
\end{align}
The leading term in the fugacity behavior is thus
\bea\label{eq:z0m}
z_0 = \left[A\frac{1}{(1-q)^s}\textrm{B}\left(s,\frac{1}{1-q}-s\right)\right]^{-1} T^{-s}
\end{align}
for $q<1$ and
\bea\label{eq:z0p}
z_0 = \left[A\frac{1}{(q-1)^s}\textrm{B}\left(s,\frac{1}{q-1}+1\right)\right]^{-1} T^{-s}
\end{align}
for $q>1$.

Simple manipulations yield the classical limits \cite{Rovenchak:2018LTP}:
\bea\label{eq:E-class}
E=\frac{s}{q(s+1)-s}NT.
\end{align}
and
\bea\label{eq:CV-class}
C_V = \left(\frac{\partial E}{\partial T}\right)_V = \frac{s}{q(s+1)-s}N.
\end{align}
for both $q<1$ and $q>1$. As expected, the case of $q=1$ corresponds to $E=sNT$.

Adding more terms in the expansions for the number of particles (\ref{eq:Nz0}) and energy (\ref{eq:Ez0}), we can obtain corrections to the classical limits for fugacity, energy, and specific heat.

For $q<1$, from (\ref{eq:Nz0}) we have
\bea\label{eq:Nz1}
N \simeq
zNAT^s \frac{1}{(1-q)^s}
\sum_{k=0}(-1)^{k}z^{k}\,\textrm{B}\left(s,\frac{k}{1-q}-s\right)
\end{align}
and
\bea\label{eq:Ez1}
E \simeq
zNAT^{s+1} \frac{1}{(1-q)^{s+1}}
\sum_{k=0}(-1)^{k}z^{k}\,
\textrm{B}\left(s+1,\frac{k}{1-q}-s-1\right).
\end{align}

The expression for the fugacity up to two terms is
\bea
z = z_0+\Delta z = z_0 + z_0^2 \,
\frac{\textrm{B}\left(s,\frac{2}{1-q}-s\right)}%
{\textrm{B}\left(s,\frac{1}{1-q}-s\right)},
\end{align}
where the correction $\Delta z \propto T^{-2s}$, yielding energy
\bea
\frac{E}{N}&=\frac{s}{q(s+1)-s}T + \frac{AT^{s+1}}{(1-q)^{s+1}}
\Delta z \,\textrm{B}\left(s+1,\frac{q}{1-q}-s\right) \nonumber\\[6pt]
&=\frac{s}{q(s+1)-s}T +\textrm{const}\,T^{1-s},\qquad
\textrm{where \ const}>0.
\end{align}
The high-temperature correction to the classical specific heat limit (\ref{eq:CV-class}) is thus negative (since $s>1$) and proportional to $T^{-s}$. The case of $q>1$ can be analyzed in a similar way [note the differences between expressions (\ref{eq:z0m}) and (\ref{eq:z0p}) for $z_0$] with the same temperature dependence obtained.

\section{Limit of $T\to 0$ for the second model}
The low-temperature limits of the first model, with the definition involving the chemical potential, are very similar to those of the ordinary Fermi-gas. In particular, it is straightforward to show that at $T=0$ the chemical potential achieves the value
\bea\label{eq:mu0}
\mu_0 \equiv \mu\Big|_{T=0} = \left(\frac{s}{A}\right)^{1/s}
\end{align}
independent of $q$. 

The second model involving fugacity, on the other hand, reveals much more interesting properties at $T\to0$. They are studied in detail in the reminder of this section.

The behavior of the fugacity $z$ in the low-temperature limit is obtained as follows. The integral in the expression defining the fugacity 
\bea\label{eq:fug0}
AT^s\intl_0^\infty \frac{x^{s-1}\,dx}{z^{-1}e_q^x+1} = 1
\end{align}
can be split into two parts,
\bea
\intl_0^\infty \frac{x^{s-1}\,dx}{z^{-1}e_q^x+1}=
\intl_0^{x_1} \frac{x^{s-1}\,dx}{z^{-1}e_q^x+1}+
\intl_{x_1}^\infty \frac{x^{s-1}\,dx}{z^{-1}e_q^x+1},
\end{align}
where $e_q^{x_1} = z$ or equivalently, $x_1 = \ln_q z$ with the $q$-logarithm defined as
\bea
&\ln_q x \equiv \frac{x^{1 - q} - 1}{1-q},
\qquad\textrm{yielding in particular}\quad
\ln_1 x = \ln x.
\end{align}
In view of different convergence radii, the two integrals are expanded into the following series:
\bea
&\intl_0^{x_1} dx\,x^{s-1}
 \Big[1+z^{-1}e_q^x\Big]^{-1}+
\intl_{x_1}^\infty dx\,x^{s-1} z(e_q^x)^{-1}
\Big[1+z(e_q^x)^{-1}\Big]^{-1}\nonumber\\[6pt]
&=
\intl_0^{x_1} dx\,x^{s-1} 
\sum_{k=0}^\infty (-1)^k \frac{1}{z^k} \left(e_q^x\right)^k
%\nonumber\\
%&
+\intl_{x_1}^\infty dx\,x^{s-1} 
\sum_{k=1}^\infty (-1)^{k+1} z^k\left(e_q^x\right)^{-k}.
\end{align}

In the limit of $T\to0$, the fugacity tends to infinity, and so does $x_1$. We thus are interested in the leading contributions from the respective integrals,
\bea
\intl_0^{x_1} dx\,x^{s-1} \left(e_q^x\right)^k = 
\frac{(1-q)^{s+\frac{k}{1-q}}}{k+(1-q)s}\,x_1^{s+\frac{k}{1-q}}
+ \ldots,\\[6pt]
\intl_{x_1}^\infty dx\,x^{s-1} \left(e_q^x\right)^{-k} = 
\frac{(1-q)^{s+\frac{k}{q-1}}}{k+(q-1)s}\,x_1^{s+\frac{k}{q-1}}
+ \ldots\,.
\end{align} 
Taking into account that for large $z$ the $q$-logarithm reduces to the power function,
\bea
x_1 = \ln_q z \simeq \frac{z^{1-q}}{1-q},
\end{align}
we finally obtain from (\ref{eq:fug0}):
\bea
A\left(T\ln_q z\right)^s (1-q)
\left[
\sum_{k=0}^{\infty}\frac{(-1)^k}{k+(1-q)s}-
\sum_{k=1}^{\infty}\frac{(-1)^k}{k+(q-1)s}
\right]
=1.
\end{align}

This expression can be rewritten using the so called Lerch transcendent $\Phi(a;b;c) $ being a generalization of Riemann's and Hurtwitz zeta-functions:
\bea
\Phi(a;b;c) = \sum_{k=0}^{\infty} \frac{a^k}{(b+c)^k},
\end{align}
namely
\bea
\lim_{T\to 0}
&T\ln_q z =
\left\{
A(1-q)
\Big[
\Phi\big(-1;1;(1-q)s\big) + \Phi\big(-1,1,1+(q-1)s\big)
\Big]
\right\}^{-1/s}.
\end{align}
Note that in the limit of $q\to 1-0$
\bea
\Phi\big(-1;1;(1-q)s\big) + \Phi\big(-1,1,1+(q-1)s\big) =
\frac{1}{(1-q)s}+\ldots,
\end{align}
where only the leading term is written, so at $T=0$
\bea
\mu = T\ln z = (s/A)^{1/s},
\end{align}
which coincides with expression (\ref{eq:mu0}) for the Fermi energy.

Denoting 
\bea
\phi(q,s) = (1-q)
\Big[
\Phi\big(-1;1;(1-q)s\big) + \Phi\big(-1,1,1+(q-1)s\big)
\Big],
\end{align}
so that
\bea\label{eq:zto0T}
\lim_{T\to 0}
&T\ln_q z = [A\phi(q,s)]^{-1/s},
\end{align}
we can also write immediately the low-temperature limit for energy $E$, where the integral contains $x^{s}$ instead of $x^{s-1}$ for $N$. Thus,
\bea
E_0 \equiv \lim_{T\to 0} E = NA (T\ln_q z)^{s+1} \phi(q,s+1) = 
N \frac{\phi(q,s+1)}{[\phi(q,s)]^{1/s+1}}.
\end{align}
After some lengthy derivations one can show that the subsequent term is linear in temperature,
\bea\label{eq:Eto0T}
E\Big|_{T\to0} =
E_0 + T\alpha(q,s)+\mathcal{O}(T)^2,
\end{align}
where $\alpha(q,s)$ is a complex expression satisfying $\alpha(1,s)=0$. The deviation of $\alpha(q,s)$ from zero is linked to the violation of the relation $e^{-x} = [e^x]^{-1}$ for the Tsallis $q$-exponentials, $e_q^{-x} \neq [e_q^x]^{-1}$, cf. the analysis of the degenerate Fermi gas in \cite[Chap.~V]{Landau&Lifshitz:1980}.

As a consequence of (\ref{eq:Eto0T}) we have an obvious result,
\bea
C_V\Big|_{T\to 0} = \alpha(q,s) \neq 0
\qquad\textrm{for}\quad q<1.
\end{align}
Certainly, this violates the third law of thermodynamics. However, non-zero heat capacities at absolute zero are known for non-equilibrium systems as well as in glasses \cite{Mauro_etal:2010} or some magnetic systems \cite{Berlie_etal:2015}.

In the limit of $T\to0$, we have for fugacity from (\ref{eq:zto0T})
\bea
z=e_q^{\frac{1}{T}[A\phi(q,s)]^{-1/s}},
\end{align}
so the occupation numbers after simple transformations become 
\bea
n_2(\eps,z,T)=
\frac{1}{\left(\frac{\eps}{[A\phi(q,s)]^{-1/s}}\right)^{1/(1-q)}+1}
\end{align}
demonstrating thus a smooth step, unlike the ordinary Fermi-gas, which is obtained from the above expression in the limit of $q\to 0$.

For $q>1$, fugacity becomes infinite as some temperature $T_0$ yielding zero heat capacity as $T_0$. To obtain $T_0$, consider the relation
\bea
N = NAT_0^s \intl_0^\infty \frac{x^{s-1}\,dx}{z^{-1}e_q^x+1} = 
NAT_0^s \intl_0^{x_0} 
\frac{x^{s-1}\,dx}{z^{-1}[1+(1-q)x]^{1/(1-q)}+1}.
\end{align}
The finite upper limit of integration is obtained from the condition
$1+(1-q)x_0 = 0$, see definition (\ref{eq:qexpTsallis-def}). So, for $z\to\infty$,
\bea
T_0 = (q-1)\left(\frac{s}{A}\right)^{1/s}.
\end{align}
Thus, in the case of $q>1$ there exists a finite non-zero minimal temperature $T_0$ in the model. It becomes zero in the limit of $q\to 1$, as expected.

\begin{figure}[h]
\centerline{%
\includegraphics[width=0.8\textwidth]{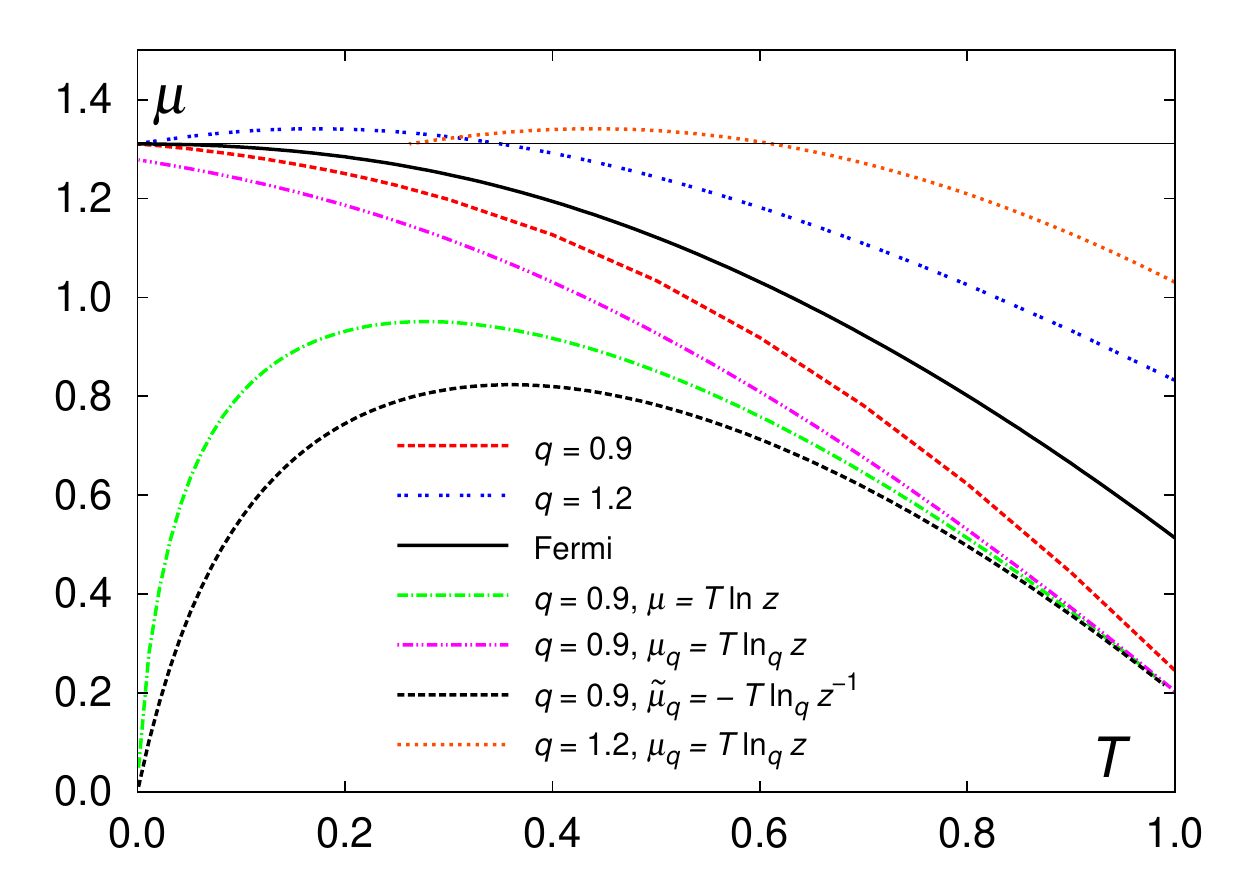}
}
\vspace*{-2ex}
\caption{Various definitions of the chemical potential in the second model as a function of temperature $T$ compared to the chemical potential in the first model and in the ordinary Fermi-gas.}
\label{fig:muq}
\end{figure}

To complete this section, we consider what would be the appropriate definition of the chemical potential in the second model, namely how it is connected with fugacity. In our derivations, the relation $T\ln_q z$ occurs naturally, which corresponds to $\mu=T\ln z$. We thus can define the quantity
\bea\label{eq:muq=Tlnqz}
\mu_q = T\ln_q z
\end{align}
as an equivalent of the chemical potential. On the other hand, from $z^{-1} = e^{\mu/T}$, substituting the ordinary exponential with the Tsallis one, we arrive at
\bea
\tilde\mu_q = -T\ln_q z^{-1}.
\end{align}
In Figure~\ref{fig:muq}, different definitions of the chemical potentials in the second model are shown in comparison with the chemical potential in the first model. As one can see, the first suggested definitions (\ref{eq:muq=Tlnqz}) has the behavior most similar to ordinary chemical potential.

\section{Discussion}
We have analyzed two phenomenological approaches to the generalization of the Fermi-distribution using the nonadditive Tsallis $q$-exponential. The first model obtained by substitution of the Gibbs factor $e^{(\eps-\mu)/T}\to e_q^{(\eps-\mu)/T}$ demonstrates the behavior at high temperatures, which significantly deviates from the classical limit. Namely, the heat capacity is not constant but instead grows infinitely for $q<1$ or tends to zero for $q>1$. The low-temperature limits of this model, on the other hand, closely resemble those of the ordinary ideal Fermi-gas.

The second model was obtained by substituting $e^{(\eps-\mu)/T}=\linebreak z^{-1}e^{\eps/T}\to z^{-1}e_q^{\eps/T}$. Note that properties of the Tsallis $q$-exponential do not allow factorization $e_q^{(\eps-\mu)/T}\neq e_q^{-\mu/T}e_q^{\eps/T}$. The differences with the first model appear fundamental. In particular, at high temperatures the specific heat tends to a constant value. The most interesting results, however, are obtained in the low-temperature domain. 

For $q<1$ in the second model, the isochoric specific heat tends to a positive constant value at $T=0$. This property might be useful in modeling magnetic systems \cite{Berlie_etal:2015} or non-ergodic systems like glasses \cite{Mauro_etal:2010}. Moreover, extrapolation of the $C_V$ curve to achieve zero value would mean negative absolute temperatures, which appeared recently in some cosmological models \cite{Szolgyen&Kocsis:2018,Cvetic_etal:2018} but are not limited to those \cite{Rapp:2012,Abraham&Penrose:2017}.

A finite (non-zero) minimal temperature obtained for $q>1$ in the second model is another exotic result though found in the literature \cite{Chung&Hassanabadi:2018}. Such behavior might be used in effective description of systems with so-called minimal momentum \cite{Fityo_etal:2008,Stetsko:2013}.

Prospects of the presented research include studies of interacting systems and analysis of other types of deformed exponentials, like the Kaniadakis $\kappa$-exponential \cite{Kaniadakis:2001,Ourabah&Tribeche:2014}, or some other types of statistics deformation \cite{Algin_etal:2015,Gavrilik_etal:2018}.

\section*{Acknowledgment}
We are grateful to Dr. Volodymyr Pastukhov and Yuri Krynytskyi for discussions.

This work was partly supported by Project FF-83F (No. 0119U002203) from the Ministry of Education and Science of Ukraine.

\bibliographystyle{unsrtnat}
%\bibliographystyle{plainnat}
%\bibliography{../full,q-Bose}

\end{document}